\documentclass[twocolumn,showpacs,amsmath,amssymb]{revtex4}
\input epsf
\usepackage{appendix}
\usepackage{rotating}
\begin{document}
\title {The dispersion relation and excitation character of a two component 
Bose Einstein Condensate}
\author{Christopher Ticknor}
\affiliation{Theoretical Division, Los Alamos National Laboratory, Los Alamos, New Mexico 87545, USA}
\date{\today}
\begin{abstract}
We present a study for the dispersion relation and character of the excitations
of a single and two component Bose Einstein Condensate (BEC).  
We study the single component dispersion for a finite BEC system and
look at examples of quasiparticles to understand and characterize the 
dispersion relation. Next we present the dispersion relation for a 
two component BEC in both the miscible and immiscible parameter regimes. 
Then we present examples of the quasiparticles for both regimes.
\end{abstract} 
\maketitle
\section{introduction}
Ultracold atomic gases have proven to be incredibly successful for studying idealized
quantum systems because essentially every aspect of the system can be tuned \cite{rmp}.
Of interest here are multicomponent quantum gases \cite{multi}.
These systems offer interacting quantum fields which can have their interactions tuned
to create a wide variety of physical scenarios. 
For example, they have been use to study
non-equilibrium excitations and motion of BECs \cite{mertes,russ}
and the miscible/immiscible transition \cite{papp,tojo,nicklas}.
Other theoretical work has been devoted to the Kibble-Zurek mechanism where
time dependent changes in the interactions lead to domain formation \cite{kz}.
The interface in an imscible two component Bose Einstein Condensate (BEC) has been studied 
in simulations where Rayleigh-Taylor instabilities have been predicted \cite{int1,int2,int3} 
and interface dynamics and excitations have been characterized \cite{int4,int5,int6}.
Additionally, these systems can be used to study subtleties in quantum field theory,
such as Goldstone modes \cite{goldstone}.

Much work has relied on the analytic dispersion relation, $\omega(k)$, for a  BEC \cite{smith},
where $\omega$ is the energy of an excitation which is characterized by wavenumber $k$.
This dispersion relation is used to understand the character of wave motion.
For superfluids, the dispersion relation can be used to estimate the breakdown of 
superfluidity as an object moves through the BEC.
This has been predicted for ultracold atomic gases \cite{vel}
and it was observed experimentally \cite{arizona}.
Additionally, it has been applied recently to dipolar BECs \cite{ryan,aniso}.

But little work has been devoted to finite quantum systems and their dispersion relation. 
For a finite  BEC, careful analysis of the dispersion relation for trapped 
dipolar BECs has been done when there is a roton-like excitation, a local minimum in 
$\omega(k)$ \cite{ryan,blair,bisset}.  Such analysis has not been done for a two component 
trapped quantum gas; in this paper we do just that. We look at the dispersion relation of a 
two component quantum.  We further characterize excitations throughout the dispersion relation
so we can understand its structure. 

The rest of this paper is broken into four sections. 
First, we review the equations of motion and formalism used to describe the BEC
at its excitations.
Second, we look at the dispersion relation of single component BEC and we
look at typical examples of excitations to understand the structure of the quantum gas. 
Third, armed with this knowledge, we will look at the two component dispersion 
relation for both a miscible and immiscible gas.
We look at examples of quasiparticle and classify them to further understand the structure 
of the dispersion relation.
Here we further previous work \cite{ct2BEC} and in particular the dispersion 
relation for the two component gas.
Finally, we remark on some conlcusions from this work.

\section{Equations of motion}    
In this work, we will study a quasi-2 dimensional (q2D) BEC.  In this set up there 
s a tight trapping axis, taken to be $z$, such that excitations in this direction are 
energetically frozen out. But the scattering is still 3D in nature. 
Furthermore, spectacular advances in imaging of q2D BEC systems have lead to single 
atom resolution \cite{microscope} and {\it in situ} imaging of a thermal phase 
transition \cite{chin}.  Additionally, such imaging has been used to observe two 
component domain formation \cite{parker}. 
 
To start with our study of a two component BEC, we break the second quantized bosonic 
field operator
into a condensate and thermal component:
$\hat\Psi_i=[\sqrt{N_i^0}\phi_i(x)+\hat\theta_i(x)]$ where $i$ labels the component 
and we have replaced $\hat a_i^0\rightarrow\sqrt{N_i^0}$.
$\hat\theta$ is the thermal operator, and it follows standard bosonic commutation relations:
$[\hat\theta_i,\hat\theta_j^*]=\delta(x-x^\prime)\delta_{ij}$.
Here $x$ represents all required coordinates, for the 2D case it will 
be $\vec\rho$.  The Gross Pitaevskii equation (GPE) governs the evolution of the 
condensate:
\begin{eqnarray}\label{gpeq} 
&&H^{GP}_k\phi_k=\left(H_k-\mu_k+D^0_k \right)\phi_k=0
\end{eqnarray}
where $H_k$ is the kinetic and potential energy for the $k^{th}$ component, 
$D^0_k$ is the direct interaction and is $
\sum_i\int dx^\prime V_{ik}(x-x^\prime)[n_i^0(x^\prime) ]$
where $n_i^0$ the condensate density, $N_i^0|\phi_i|^2$ and $\phi_i$ is unit normalized.

We will use the Bogoliubov transformation \cite{fetter,pu,eddy} to describe the thermal 
part: $\hat\theta_i(x)=\sum_{\gamma} [u_{i\gamma}(x)\hat a_{\gamma} e^{-i\omega_\gamma t}
+v_{i\gamma}^*(x)\hat a^*_{\gamma} e^{i\omega_\gamma t}]$ where 
$\hat a_{\gamma}$ ($\hat a^*_{\gamma}$) is 
the bosonic annihilation (creation) operators for the mode $\gamma$ with projection on 
to the $k^{th}$ component.  These operators obey the commutation relations: 
$[\hat a_{\gamma},\hat a_{\beta}^*]=\delta_{\gamma\beta}$ and  
$[\hat a^*_{\gamma},\hat a^*_{\beta}]=[\hat a_{\gamma},\hat a_{\beta}]=0$.
This excitation has a spatial distribution given by 
the quasi-particle $u_{i\gamma}$ and the hole $v_{i\gamma}$. 
The normalization of the quasi-particle wavefunction is 
$1=\sum_i\int dx (|u_{i\gamma}(x)|^2-|v_{i\gamma}(x)|^2)$.  

The equation for Bogoliubov quasiparticles is:
\begin{eqnarray}\nonumber
&&\left( \begin{array}{cc}H^{GP}_k& 0 \\0 & H^{GP}_k\end{array} \right)
\left(\begin{array}{cc}u_{k\gamma}\\v_{k\gamma}\end{array}\right)
+\sum_i\left( \begin{array}{cc}X^0_{ik}& X^0_{ik}\\
X^0_{ik} & X^0_{ik}\end{array} \right)
\left(\begin{array}{cc}u_{i\gamma}\\v_{i\gamma}\end{array}\right)
\\&&=
\omega_\gamma\left(\begin{array}{cc}u_{k\gamma}\\-v_{k\gamma}\end{array}\right).
\label{uveq}
\end{eqnarray}
We have assumed $\phi_i$ to be real and 
$X^0_{ik}=\int dx^\prime V_{ik}(x-x^\prime)[n_{ik}^0(x,x^\prime)]$ where $n_{ij}^0$ is the
condensate  correlation function $\sqrt{N_i^0N_k^0}\phi_i(x)\phi_k(x^\prime)$.

For this study q2D gas we use $V_{ij}(\vec \rho-\vec\rho^\prime)=g_{ij}\delta(\vec\rho-\vec\rho^\prime)$.
For this system
$g_{ij}=N\sqrt{8\pi} \hbar^2a_{ij}/m l_z$ is the strength of the contact 
interaction, $a_{ij}$ is the 3D s-wave scattering length between components $i$ and $j$
($a_{ij}\ll l_z$), $l_z=\sqrt{\hbar/m \omega_z }$ is the axial harmonic 
oscillator length and $\omega_z$ is the tight trapping frequency.  
Throughout this work we will rescale the equations into trap units, 
so the energy scale is $\hbar\omega_\rho$ and the length scale is 
$l_\rho=\sqrt{\hbar/m \omega_\rho }$ where $\omega_\rho$ is the trapping frequency in 
$x-y$ plane.

For the rest of the paper we focus on three examples.  
First we look at a single component system with $g=1000$ with $\mu=17.9\hbar\omega_\rho$.
The next two examples have
$g=500$ with $g_{11}=0.99g$, $g_{22}=1.01g$, and $g_{12}=0.5g$ $(=g_{21})$ for the miscible example
or $g_{12}=2g$ for the immiscible example.
The chemical potentials are $\mu=16.6\hbar\omega_\rho$ for the miscible example
and $\mu=18.05\hbar\omega_\rho$.
The two chemical potentials of each component are not equal but they are within a percent 
of each other.  The miscible ground state has two near identical ground states.  In contrast, 
the immiscible ground state is a state with broken symmetry and one condensate is
on the left while the other is on the right. 

We can estimate what these examples represent in terms of experiments, 
say $g$=500 we pick $N_i=10^4$, $\omega_z$/$\omega_\rho$=25, and $a_{ii}$=100.
Then this example corresponds to radial trapping frequencies, 
of $2\pi\times$38 Hz for K and $2\pi\times$11 Hz for Cs.
Further details of how we solve these equations (\ref{gpeq}) and (\ref{uveq}) 
for this system appear in Ref. \cite{ctHFB,ct2BEC}.

\section{excitations and dispersion relations of a single BEC}
A common and analytic dispersion relation is obtained by assuming a plane 
wave excitation, $e^{-ikx}$, perturbation of a uniform condensate \cite{smith}. This leads 
to a dispersion relation:
\begin{eqnarray}
\omega(k)=\sqrt{\epsilon^2+2\mu \epsilon}
 \label{pwBDG}
\end{eqnarray}
where $\epsilon=k^2/2$ is the free particle kinetic energy.
The speed of sound of is $c=\sqrt{\mu}$ and is defined from the small $k$ behavior of the 
relation: $\omega\sim c k$.
The Bogoliubov dispersion relation was generalized for two component BEC in Ref. \cite{eddy}.
That work showed the dispersion relation has two branches, in-phase 
excitations (higher in energy) and out-of-phase excitations (lower in energy for given $k$).
For a finite system, whose spectrum is discrete,  we use the root mean squared of $k$ 
for each excitation to construct a dispersion relation:
\begin{eqnarray}
k=\left({\sum_i\int d\vec k k^2(|u_{i\alpha}(\vec k)|^2+|v_{i\alpha}(\vec k)|^2)\over
\sum_i\int d\vec k(|u_{i\alpha}(\vec k)|^2+|v_{i\alpha}(\vec k)|^2)}\right)^{1/2}. 
\label{kdef}
\end{eqnarray}
This leads to a scatter plot of $\omega_\gamma(k)$. 
This has been done to study the momentum dependence of roton like excitations  
in dipolar BECs \cite{ryan,blair,bisset}.

To further classify the modes and understand the dispersion relation we can rewrite 
the quasiparticles into cylindrical coordinates
as $u_\alpha(\rho,\phi)=f_\alpha(\rho)e^{im\phi}$ where $m$ is the azimuthal 
quantum number and $f_\alpha$ is the radial wavefunction.
This classification of modes will help us understand the structure of the dispersion 
relation.
Such an analysis shows that there are many different types of excitations,
from phonons to surfaces.

\begin{figure}
\includegraphics[width=80mm]{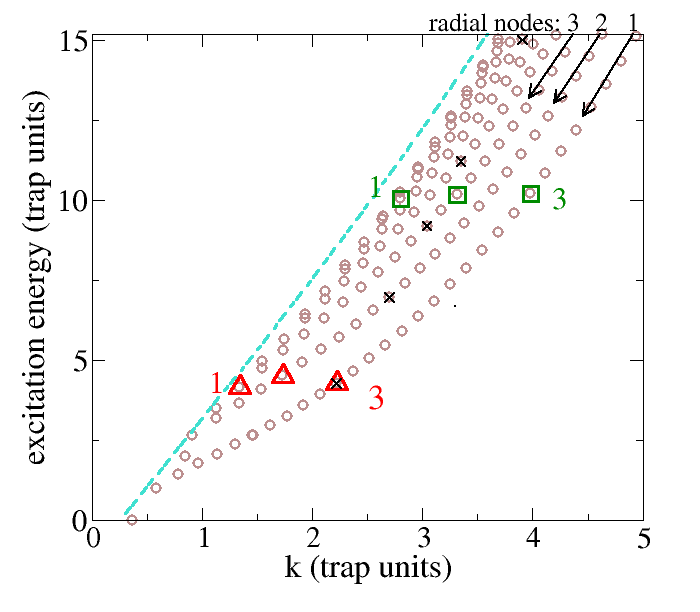} 
\caption{(Color online)  The dispersion relation for a single component gas.
There is a clear structure and the modes are readily classified, see text. 
The quasiparticles highlighted by red triangles are plotted in Fig. \ref{qp0} (a)
and green squares are plotted in Fig. \ref{qp0} (b).
The turquoise dashed line is $\omega=\sqrt{\mu} (k+k_0)$ where $\mu=17.9\hbar\omega_\rho$.
The black $\times$s are modes with $m=11$ and different numbers of radial nodes.  }
\label{dsing}
\end{figure}

Fig. \ref{dsing} shows the dispersion relation for a single component BEC.
There are three arrows illustrating the structure of dispersion 
relation. The arrows move down and left along a set of excitations
that are related to each other by changing the azimuthal quantum number,
while all having the same number of radial nodes.
At the top right of the plot we have labeled the first three of these curves 
1, 2, and 3. 
To be explicit, we count the node at $\rho=0$ (for $|m|>0$) but not the zero
at large $\rho$.
When one  moves to a different curve, the number of radial nodes is changed. 
These curves are fairly distinct until they encounter the linear dispersion regime 
where they are born.  
The dashed turquoise line shows the phonon or linear behavior extracted from 
Eq. \ref{pwBDG}.  In this figure we have off set the line in $k$ to guide the eye. 

\begin{figure}
\centerline{\includegraphics[width=90mm]{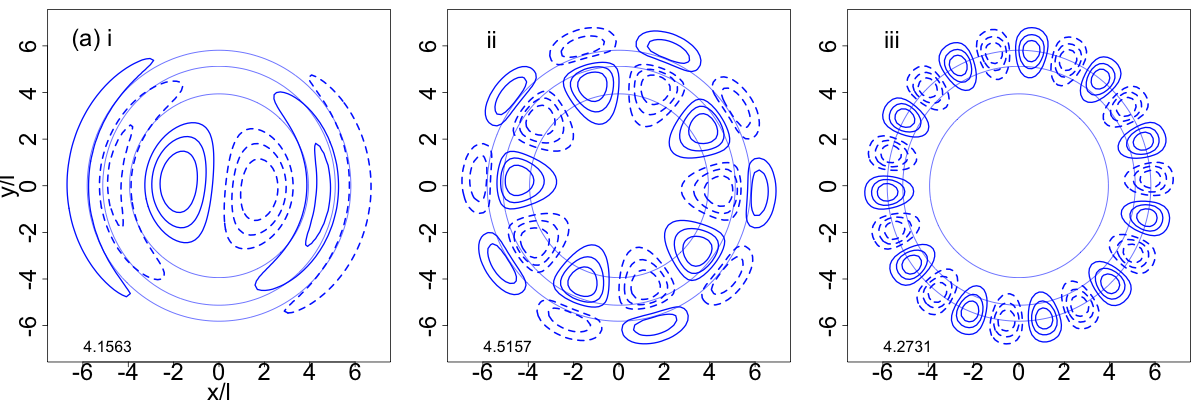}} 
\centerline{\includegraphics[width=90mm]{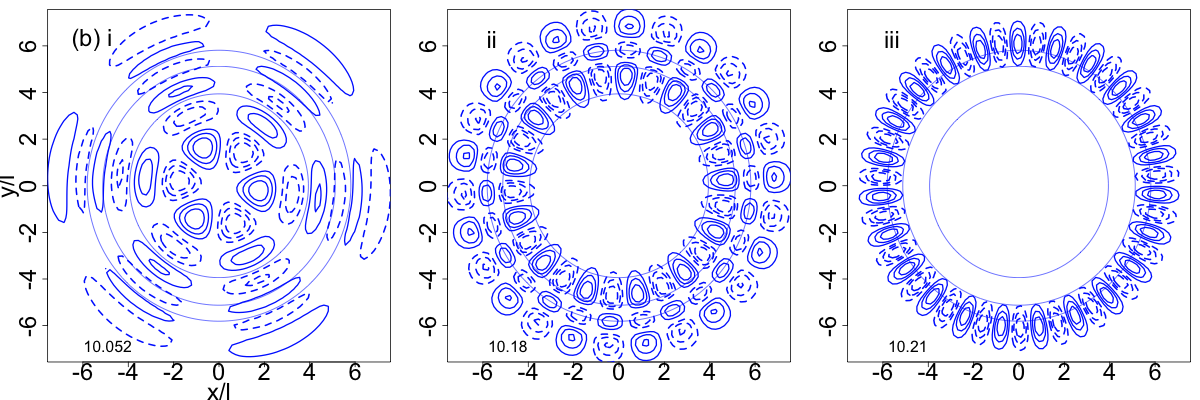}} 
\caption{(Color online) Examples of quasiparticles are show as a function of space (in trap units)
with energies near (a) 4 and  (b) 10 $\hbar\omega_\rho$.
The example quasiparticles are highlighted in Fig. \ref{dsing} with red triangles
and green squares.  The energy of the excitation is written in the lower left corner of
each plot in trap units.
}
\label{qp0}
\end{figure}

We will now illustrate these points by looking at a few quasiparticle
wavefunctions ($u_{i\alpha}$).
In Fig. \ref{qp0} we show several different quasiparticles.
In (a) three wavefunctions are shown with energy near 4$\hbar\omega_\rho$, 
highlighted with red triangles in Fig. \ref{dsing}. 
Then in (b) three wavefunctions are shown with energies near 10$\hbar\omega_\rho$, 
highlighted with green squares in Fig. \ref{dsing}.
For each plot the contours of $u_{\alpha}$ are shown. 
We plot solid contours at 1/4, 1/2, and 3/4 of the maximum value of
wavefunction. We also plot dashed contours of  -1/4, -1/2, and -3/4 the maximum value.
This shows that as this excitation evolves in time, the density moves from the 
region bounded by the dashed curves to the regions bounded by solid curves
and back again with a frequency of $\omega_\alpha$.
The condensate wave function is also shown as a faint line in the background of the plot
with contours 1/4, 1/2, and 3/4 of the condensates maximum value.
We will now discuss each wavefunction shown in Fig. \ref{qp0}.

Fig. \ref{qp0} (a) shows three different quasiparticles with energy near
4$\hbar\omega_\rho$. These are highlighted in Fig. \ref{dsing} with red triangles. 
In Fig. \ref{qp0} (a) i, a quasiparticle with a low value of $k$ is shown.  
It is more phonon like being near the linear part of the dispersion.   
In this case it has a large amplitude near the center of the condensate. 
Additionally it has 3 radial nodes and $m$=1.
In  Fig. \ref{qp0} (a) ii, we show a quasiparticle with moderate $k$.
This mode is characterized by $m$=5 and 2 radial nodes.
Finally in  Fig. \ref{qp0} (a) iii, we show a surface mode with large $k$. 
This excitation has 1 radial node (at 0) and $m$=11. This mode has no amplitude 
at the center of the condensate.

These three modes illustrate that for an excitation energy held roughly
constant, there are several different excitations that can occur.  The lowest $k$
mode is phonon like, then at largest $k$ mode is a surface excitation.
As one moves between these two extremes, the excitation moves from the center of the 
trap toward the surface. This happens by the wavefunction losing radial nodes and   
increasing the azimuthal quantum number at roughly a constant energy.

One other way to look at the structure of the dispersion relation is to freeze $m$
and vary the number of radial nodes. We have done this for one example with $m=11$, 
in Fig. \ref{dsing} these energies are marked with black $\times$s.  They start with 
large $k$ and lower energy.  Then as a radial node is added, the $\times$ moves up in 
energy and to slightly higher $k$.

We now move to study the quasiparticles with higher energy ($\sim10\hbar\omega_\rho$), 
these are shown in Fig. \ref{qp0} (b). 
In Fig. \ref{qp0} (b) i, we show a low $k$ mode which is phonon like.  
This mode has a large amplitude near the center of the trap and the radial wavefunction 
has 6 nodes and $m$=3.
This mode is similar to the mode shown in (a) ii, in that it has a large amplitude
near the center of the trap with a relatively small $m$ (for its energy).  

Fig. \ref{qp0} (b) ii shows a mode with moderate $k$ and has 3 radial nodes
and $m$=13.  
It is worth noting that if we follow the curve which contains the modes with 3 
radial nodes, we find that it contains the mode shown in (a) i.
Finally, in Fig. \ref{qp0} (b) iii we show a mode with large $k$. This excitation is a 
surface mode with $m$=23 and one radial node at 0. 
Now that we have characterized the dispersion relation by looking at the 
particular excitation modes, we move on to look at the dispersion
relation and excitations of the 2 component BEC.

\begin{figure}
\includegraphics[width=80mm]{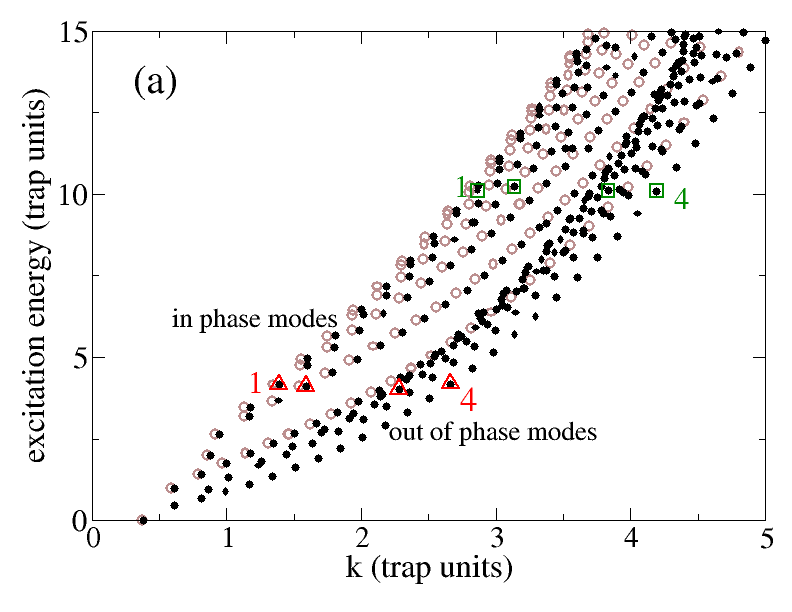} 
\includegraphics[width=80mm]{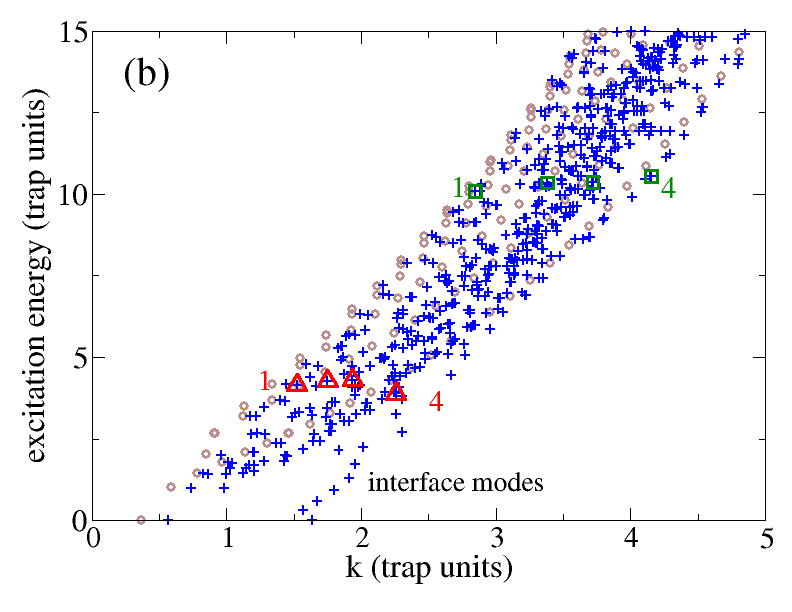} 
\caption{(Color online) (a)  The dispersion relation for a miscible system is shown 
(black dots). (b) The dispersion relation for an immiscible system (blue +) is shown.
The single component dispersion relation is show as brown circles for both (a) and (b).
}
\label{dis}
\end{figure}
\section{excitations and dispersion relations of a two component BEC}

Comparing the one and two component dispersion relations we see that
the primary difference is that another quantum 
number is required to tell us the ``spin'' state of the excitation, 
this leads to twice as many excitations. 
If we classify the modes in cylindrical coordinates
the wavefunction now reads $u_{i\alpha}(\rho,\phi)=f_\alpha(\rho)e^{im\phi}\chi_i$
where $\chi_i$ is the spin state of the excitation.
This extra index is of course from the fact that we have gone from a one by one matrix 
descrbing spin to a two by two matrix.
In the miscible example the modes are classified by the 
relative motion of the components: in-phase and out-of-phase modes
excitations.  Therefore spin eigenstates look like
$\chi_{\pm}=\chi_1\pm\chi_2$.  This is not rigorously true because of the 
different interaction strengths, $g_{ii}$.

In Fig. \ref{dis} (a) we show the dispersion curve of the miscible example
(black dots) and the single component BEC (brown circles).
The 2 component dispersion relation is readily broken into two regimes, particularly at low
energy: in-phase and out-of-phase modes.
Comparing the single component and 2 component dispersion relations, we see that
in-phase modes line up well with single component dispersion.
Meaning when components 1 and 2 move together, $\chi_+$, they strongly 
resemble the single component modes. Therefore these modes are then readily classified.
But lower in energy there is another manifold of excitations. These are out of
phase excitations, where  components 1 and 2 move opposite to one another, $\chi_-$.  These modes
can still be easily classified based on their azimuthal symmetry and the number of 
radial nodes.

For the immiscible system, the dispersion relation is shown in Fig. \ref{dis} (b).  
It is just a mess, this is because there is no longer a symmetry to 
protect the modes from hybridizing, and it is much harder to classify modes.  
It is worth noting that the dispersion relation has 
roughly the same extent in $k$ as the single component dispersion relation. 
An interesting addition to this dispersion relation is a set of low energy, large $k$ 
modes which are localized at the interface, these are labeled in Fig. \ref{dis} (b). 

Now that we have examined dispersion relation for the two component system, we will 
look at several examples of quasiparticles for the miscible and immiscible systems
to further understand the structure of their dispersion relations.

\begin{figure*}
\includegraphics[width=180mm]{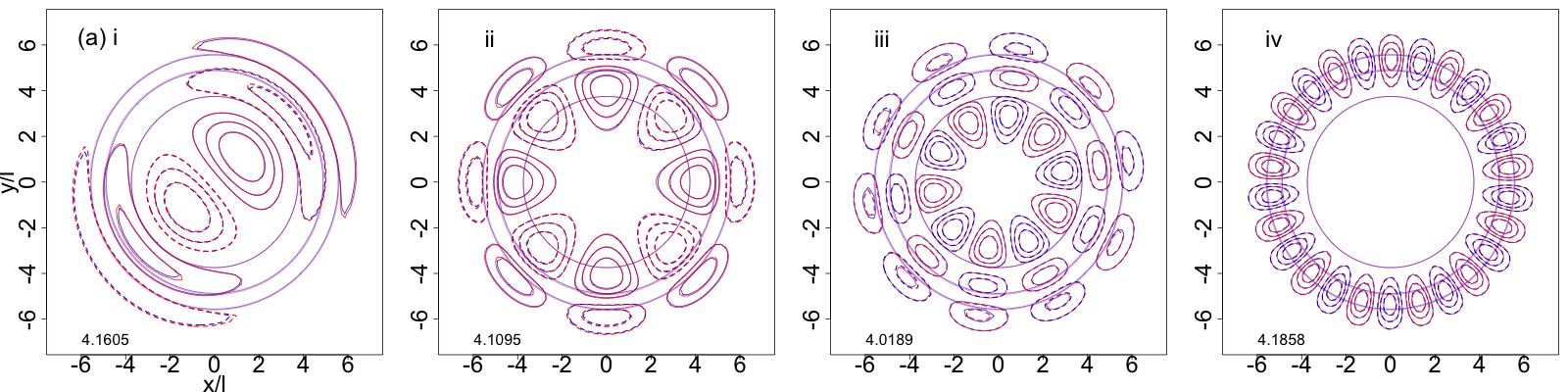}  %
\includegraphics[width=180mm]{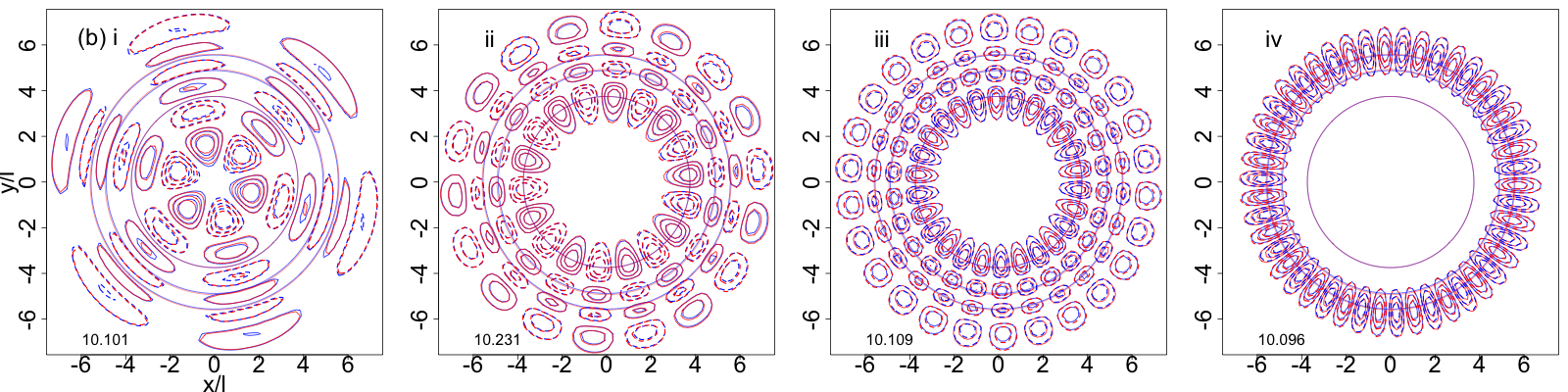} %
\caption{(Color online) Quasiparticles for miscible system are shown as a function of
space (in trap units).
(a) The quasiparticles with energies near 4$\hbar\omega_\rho$ are shown and
(b) the quasiparticles with energies near 10$\hbar\omega_\rho$ are shown. 
For each set of quasiparticles, they start at small $k$ (left, i) and move to  
to large $k$ (right, iv).  These modes are highlighted by either red triangles 
or green squares in Fig. \ref{dis} (a).
The energy of the excitation is written in the lower left corner of
each plot in trap units.
}\label{qp1}
\end{figure*}

Fig. \ref{qp1} shows several quasiparticles for the miscible example near 
two different energies of (a) 4 and (b) 10$\hbar\omega_\rho$.
In the case where we have two components, we have colored one component
blue and the other red; this coloring is shared by the quasiparticles. 
The dashed contours have values of: -0.75, -0.5, and -0.25 the maximum value of $u_{1\alpha}$,
and solid contours have values of: 0.25. 0.5, and 0.75 the maximum value of 
$|u_{1\alpha}(\vec \rho)|$.

Fig. \ref{qp1} (a) i shows an in-phase, $m$=1 phonon like excitation with 3 radial nodes.
The solid (dashed) lines of component 1 overlap the solid (dashed) lines of component 2.
This mode is very similar to the one in Fig \ref{qp0} (a) i. 
Fig. \ref{qp1} (a) ii shows an excitation with radial 2 nodes, $m$=4, and the components in-phase.
These two modes are both in-phase modes and strongly resemble modes in the 
single component system.  Fig. \ref{qp1} (a) iii shows an excitation with radial 3 nodes 
and $m$=5, but for this mode the components move out-of-phase.  
The solid (dashed) lines of component 1 overlap the dashed (solid) lines of component 2.
In this case, we are in the lower manifold of the dispersion relation, which
is essentially all out-of-phase modes at low energy. This mode has a large $k$ value for
its energy.  The similar in-phase mode, has much higher energy.
In Fig. \ref{qp1} (a) iv an out-of-phase surface mode is shown.  It has  one radial 
node at 0 and $m$=13.  At a given energy, this surface mode has the largest $k$. 

Fig. \ref{qp1} (b) i shows an in-phase, $m$=3 phonon like excitation with 6 radial 
nodes.  Again, this is very similar to the excitations in Fig \ref{qp0} (b) i. 
Fig. \ref{qp1} (b) ii shows an excitation with 4 radial nodes, $m$=9, 
and the components move in-phase.
Fig. \ref{qp1} (b) iii shows an excitation with 4 radial nodes, $m$=13, but 
the components move out-of-phase.
Fig. \ref{qp1} (b) iv shows an out-of-phase surface mode, with 1 radial 
mode (at 0) and $m$=23. 
Looking at the dispersion in Fig. \ref{dis} (b), we see that the modes in (b) iii and (b) iv 
are a regime where the 
in and out-of-phase manifolds merge, but both modes are out-of-phase modes.
 
\begin{figure*}
\includegraphics[width=180mm]{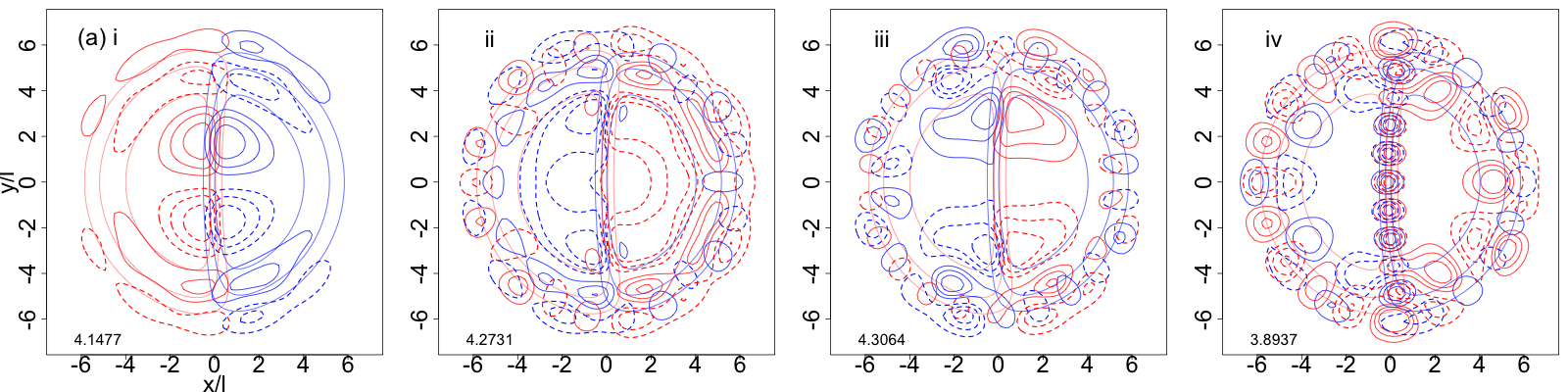} 
\includegraphics[width=180mm]{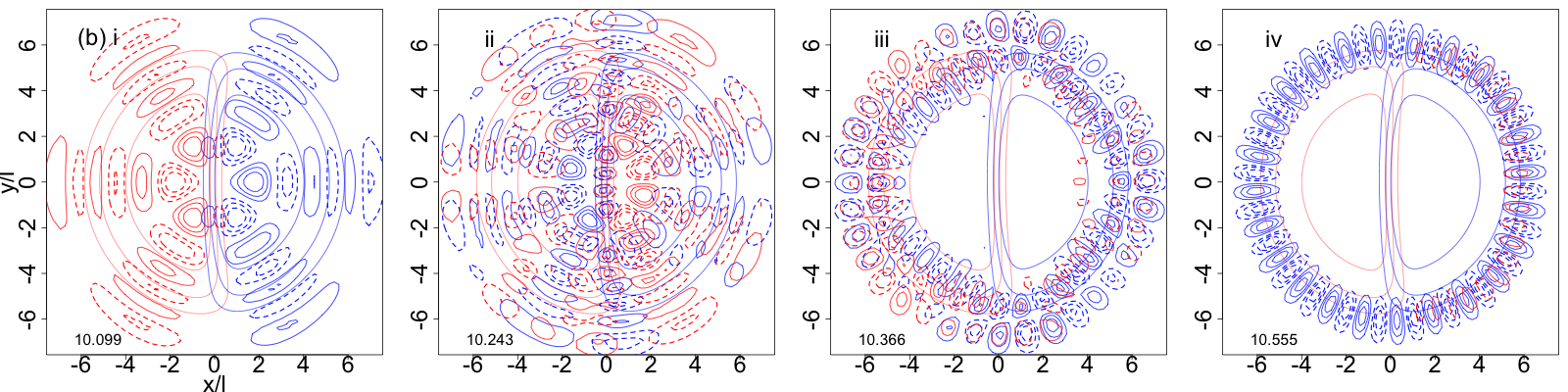}
\caption{(Color online) Quasiparticles for the immiscible system are shown
as a function of space (trap units).
(a) The quasiparticles with energies near 4$\hbar\omega_\rho$ are shown and
(b) the quasiparticles with energies near 10$\hbar\omega_\rho$ are shown. 
For each set of quasiparticles, they start at small $k$ (left, i) and move to  
to large $k$ (right, iv).  These modes are highlighted by either red triangles 
or green squares in Fig. \ref{dis} (b).
The energy of the excitation is written in the lower left corner of each plot in trap units.
}\label{qp2}
\end{figure*}

Now we will look at the quasiparticles from an immiscible system, which corresponds to
the dispersion relation in Fig. \ref{dis} (b).
Fig. \ref{qp2} shows the quasiparticles for an immiscible system at two different 
energies (a) 4 and (b) 10$\hbar\omega_\rho$.
For this example, the blue condensate is on the right and red is on the left. 

For this system, there is no clear means to classify the modes.
We now find that there are excitations that {mix} the components, 
for example the amplitude of $u_{1\alpha}$ has a large 
amplitude in condensate 2.  There are ``partisan'' excitations where the amplitude 
of an excitation overlaps strongly with its own condensate and not the other condensate.
Another prominent theme will be hybridizing of modes which resemble
modes from the miscible system.
Also some excitations are not an equal mixture of the components.  

In the dispersion curve, Fig. \ref{dis} (b), we have labeled the modes that are 
concentrated at the interface between the condensates.
These are low energy, high $k$ modes.  The modes were previously explored in Ref. \cite{ct2BEC}.

First we will look at the lower energy excitation modes, with about 4$\hbar\omega_\rho$.
These modes are highlighted by red triangles in Fig. \ref{dis} (b).
Fig. \ref{qp2} (a) i shows a mode that looks like an $m$=1 phonon with 2 radial nodes,
like those shown in Fig. \ref{qp0}(a) i or  Fig \ref{qp1}(a) i.
The excitations are partisan and localized to their condensate.
Fig. \ref{qp2} (a) ii shows a mode that is not readily characterized.
It looks like a breathing mode where excitation of 1 (2) moves from the middle to 
the outside of condensate 2 (1), meaning it is a mixing excitation where the bulk of 
the excitation is in the other condensate. There is a large component of excitation 
on the surface of the condensate.

Fig. \ref{qp2} (a) iii shows another mode that is not readily characterized.
It looks like a slosh mode along the interface, but it also mixes the components:
the excitation of 1 (2) moves back and forth along the interface in  condensate 2 (1).
Fig. \ref{qp2} (a)  iv shows an interface mode with some surface character.
This mode has large $k$ for its energy, this is characteristic of both surface 
and interface excitations.

Having looked at examples of lower energy excitations for the immiscible system, 
we will now look at higher energy excitations.  The modes we now look at are marked by 
green squares in  Fig. \ref{dis} (b).
Fig. \ref{qp2} (b) i shows a mode that looks like an $m$=3 phonon with 6 radial nodes.
The excitations are localized to their condensate, but the excitation is a collusion
between the two components to retain a familiar form like those in Figs. \ref{qp0} (b) i 
and  \ref{qp1} (b) i.
Fig. \ref{qp2} (b) ii shows a mode that looks like it has $m=4$ with many radial nodes ($\sim6$), 
but the central part of the mode mixes the components.  Additionally, there is a prominent
interface bending mode.  Generally, this is a very delocalized excitation with amplitude  
across the whole system. It is rather typical of the modes in this region of the dispersion
relation.

Fig. \ref{qp2} (b) iii shows a surface mode with roughly one radial node and $m\sim$18. 
This mode is roughly localized to its condensate, but does have some amplitude over top
the other condensate. 
Finally, Fig. \ref{qp2} (b) iv shows a surface mode which is dominated by 
component 1 (blue) that is delocalized across the whole system.
Interestingly, there is some component 2 (red) excitation, and 
its amplitude is mixing in the 1 condensate (right).
Nearby in energy there is another surface mode  which is dominated by component 2 (red) 
and looks very similar if the two components are switched. The energy difference is because
of the different values of $g_{11}$ and $g_{22}$.

\section{conclusions}

In this work we have looked at the dispersion relation of a two component BEC. 
Do to this we first looked at a dispersion relation from a single component BEC, Fig. \ref{dsing}.
We showed examples of various excitation modes, Fig \ref{qp0}.  It should be noted
such excitations will not be observed individually, rather they help us understand the dispersion relation.
Then we looked at the miscible and immiscible dispersion relationship of a two component BEC, Fig \ref{dis}.
For the miscible, we saw that there is a clear way to characterize the miscible excitation spectrum: 
in-phase and out-of-phase modes. 
Examples of miscible modes are shown in Fig. \ref{qp1}.  This figure shows
at a fixed $k$, the out-of-phase modes are generally lower in energy than in-phase modes. 

For the immiscible system where one BEC is on the left and the other on the right side of the trap, 
the dispersion relation is much harder to characterize.  
This dispersion relation shown in Fig \ref{dis} (b) shows little structure.
General characterization of modes, like surface, interface, and bulk, is still possible but hard to do in general. 
One common feature is hybrization of the modes. This is where a mode resembles one from the miscible system, 
but due to the lack of symmetry, other modes have mixed or hybridized into a single mode.
Additionally, the spin states of the immiscible excitations are much more ambiguous.  
They range from equal parts of component 1 and 2 to single component dominated.
One other means to classify a mode is the overlap of the excitation with its condensate.
There are partisan excitations which essentially only overlap with their condensate, and 
there are mixing excitations where the excitations overlap strongly with the other component's 
condensate.  Examples of such modes are shown in Fig. \ref{qp2}.

Future work will be to develop an understanding the behavior
of multicomponent quantum gases by looking at correlation functions.  In particular, it will
be important to understand the density-density correlations of on and off component 
contributions.  After this we will extend this work to multicomponent dipolar BECs \cite{RIMT}
and their coherence properties \cite{ctg2}.

\begin{acknowledgments}
The author is pleased to acknowledge excellent discussions and suggestions from R. N. Bisset and E. Chisolm.
The author gratefully acknowledges support from
LDRD ECR and LANL which is operated by LANS, LLC for the NNSA 
of the U.S. DOE under Contract No. DE-AC52-06NA25396. 
\end{acknowledgments}
\bibliographystyle{amsplain}

\end{document}